\documentclass[12pt]{article}
\pdfoutput=1
\usepackage{color,amsmath,amssymb,exscale,psfrag,epsfig}
\usepackage{cite,color,url}
\usepackage{pdfpages}
\usepackage{graphicx}
\usepackage{slashed}
\usepackage{url}
\usepackage{makecell}
\usepackage[utf8]{inputenc}
\usepackage[T1]{fontenc}
\usepackage{xcolor}
\usepackage{centernot}
\usepackage{enumerate}
\usepackage{subfig}
\usepackage{float}
\usepackage{placeins}
\usepackage{bm}
\usepackage{floatpag}
\usepackage{tcolorbox}

\usepackage[colorlinks=true
,urlcolor=blue
,anchorcolor=blue
,citecolor=blue
,filecolor=blue
,linkcolor=blue
,menucolor=blue
,linktocpage=true
,pdfproducer=medialab
]{hyperref}
\input epsf
\usepackage{lineno}

\def\bea{\begin{eqnarray}}
\def\eea{\end{eqnarray}}

\definecolor{nicered}{rgb}{0.7,0.1,0.1}
\definecolor{nicegreen}{rgb}{0.1,0.5,0.1}
\hypersetup{colorlinks,citecolor= nicegreen,linkcolor= nicered}

\def\be{\begin{equation}}
\def\te{\end{equation}}
\def\ee{\end{equation}}
\def\ba{\begin{eqnarray}}
\def\bea{\begin{eqnarray}}

\def\tea{\end{eqnarray}}
\def\ea{\end{eqnarray}}
\def\eea{\end{eqnarray}}

\def\bfra{\begin{frame}}
\def\efra{\end{frame}}
\def\al#1\fal{\begin{align}#1\end{align}}
\def\bfra#1\efra{\begin{frame}#1\end{frame}}

\catcode`\@=11
\def\lsim{\mathrel{\mathpalette\@versim<}}
\def\gsim{\mathrel{\mathpalette\@versim>}}
\def\@versim#1#2{\vcenter{\offinterlineskip
\ialign{$\m@th#1\hfil##\hfil$\crcr#2\crcr\sim\crcr } }}
\catcode`\@=12
\catcode`@=12
\begin{document}
\thispagestyle{empty}
\begin{flushright}
ICAS 051/20
\end{flushright}
\vspace{-0.2in}
\begin{center}
	{\Large \bf  COVID-19 mild cases determination from 

	correlating  COVID-line calls to reported cases  } \\
\vspace{0.2in}
{\bf Ezequiel Alvarez$^{(a)\dagger}$ and Franco Marsico$^{(b)\star}$
\vspace{0.2in} \\
{\sl $^{(a)}$ International Center for Advanced Studies (ICAS) and CONICET, UNSAM,\\ 
	Campus Miguelete, 25 de Mayo y Francia, CP1650, San Martín, Buenos Aires, Argentina }
\\[1ex]
{\sl $^{(b)}$ Ministerio de Salud de la Provincia de Buenos Aires,\\
Av. 51 1120, CP1900, La Plata, Buenos Aires, Argentina}
	}
\end{center}

\begin{minipage}[t]{.4\textwidth}
\begin{tcolorbox}[width=.98\textwidth]
{\small
	{\bf  What is already known on this topic: } There are always more COVID-19 cases than those reported.
	}
\end{tcolorbox}
\end{minipage}
\begin{minipage}[t]{.6\textwidth}
\begin{tcolorbox}[width=.98\textwidth]
{\small
        {\bf  What this study adds: } A simple algorithm to inexpensively estimate the total number of symptomatic COVID-19 cases through a correlation in COVID-19 line phone calls data.
        }

\end{tcolorbox}
\end{minipage}

\begin{abstract}

	\noindent{{\bf Background:}}
	One of the most challenging keys to understand COVID-19 evolution is to have a
measure on those mild cases which are never tested because their few symptoms are soft
and/or fade away soon. The problem is not only that they are difficult to identify and
test, but also that it is believed that they may constitute the bulk of the cases and could
be crucial in the pandemic equation.


	\noindent{{\bf Methods:}} We present a novel algorithm to extract the number of these mild cases
by correlating a COVID-line calls to reported cases in given districts. The key
assumption is to realize that, being a highly contagious disease, the number of calls
by mild cases should be proportional to the number of reported cases. Whereas a
background of calls not related to infected people should be proportional to the district
population.


	\noindent{{\bf Results:}}
	We find that for Buenos Aires Province, in addition to the background, there are in signal $6.6\pm0.4$ calls per
each reported COVID-19 case.  Using this we estimate in Buenos Aires Province $20\pm 2$ COVID-19 symptomatic cases
for each one reported.


	\noindent{{\bf Conclusions:}}
	A very simple algorithm that models the COVID-line calls as sum of signal plus
background allows to estimate the crucial number of the rate of symptomatic to reported
COVID-19 cases in a given district. The result from this method is an early and inexpensive
estimate and should be contrasted to other methods such as serology and/or massive
testing.

\end{abstract}

\vspace*{2mm}
\noindent {\footnotesize E-mail:
{\tt 
$\dagger$ \href{mailto:sequi@unsam.edu.ar}{sequi@unsam.edu.ar},
$\star$ \href{mailto:fmarsico@mincyt.gob.ar}{fmarsico@mincyt.gob.ar},
}}

\newpage

COVID-19 Pandemic is impacting on World's health and economy with an unprecedented strength \cite{coronavirus,coronavirus2,fauci}. Among the major challenges in mitigating the pandemic effect is assessing the real number of infected people at any time \cite{fauci,asym2,sang}.  This key information is not only useful for determining health policies, but also for estimating the level of immunity in society which provides a reference framework to decide the re-opening of economic and other activities.  Although the natural method for obtaining this information including mild symptom cases\footnote{Observe that our precise definition of mild case (see below) may not agree with others found in the literature.} would be to test persons upon the minimal symptom, COVID-19 is a disease with a large fraction of mild cases and therefore its cost and logistics is usually beyond the affordable.  In this work we address a novel method to estimate the total number of symptomatic infected people, including mild cases, by correlating COVID-line phone calls to lab-confirmed reported cases.  The main idea is that, since this is a highly contagious disease, then calls coming from infected people are proportional to the number of lab-confirmed people in that area and time, whereas other calls correspond to a background proportional to the population in the area.  By measuring number of calls in different scenarios, we can fit the proportionality coefficient and estimate the total number of infected people, even though a fraction of these will not reach the threshold to be derived to a laboratory diagnostic.  The idea of distinguishing a signal in a large dataset of queries is present in many schemes such as Google Flu \cite{google-flu} and others \cite{rapid-epidemic}.  However, the present method is not only considerably simpler and straightforward to be implemented in any country or region, but also is specially designed for a contagious disease with the particular feature of the mild cases such as the COVID-19.  The algorithm is based on a few reasonable hypotheses, is useful in any sub-testing scenario --as is the case in most of the countries--, and the presented general framework can be used in tackling other diseases and/or catastrophes, beyond the COVID-19 Pandemic.   The method, being statistical, does not allow to identify the mild cases, but to estimate their number.  In the following paragraphs we present the details of the algorithm, and we apply it to a real case scenario in Buenos Aires Province (PBA for its acronym in Spanish).

Along this work we model the number of calls to a COVID-line in a given district and during a given period of time by using a simple assumption of signal and background.   We define signal to those calls due to real COVID-19 cases, and background to those calls due to similar symptoms and/or other causes but that do not correspond to real infections.  It is key to clarify that the real COVID-19 cases that constitute the signal calls do not necessarily correspond to people whose symptoms will drive them to have a laboratory confirmation of their condition.  The people who has symptoms to place the COVID-call, but that their symptoms and evolution does not reach the threshold to have a laboratory test confirmation is what we call infected with mild symptoms.  As a matter of fact, we assume the compelling  hypothesis that {\it the number of signal calls is proportional to the number of lab-confirmed cases}.  On the other hand, we assume that the number of background calls is solely proportional to the district population.  This assumption is reasonable as far as the studied populations have similar social behavior and there are not major changes in the social conditions --as for instance temperature and weather-- during the analyzed period of time.  

Within the above hypothesis, we model the number of call received in district $j$ in a given period of time as
\begin{eqnarray}
	n_C^{(j)} = \theta_P \, N_P^{(j)} + \theta_I \, N_I^{(j)}.
	\label{eq:fit}
\end{eqnarray}
Where $N_P^{(j)}$ is the population and $N_I^{(j)}$ is the number of lab-confirmed infections at district $j$ and at the given period of time. $n_C^{(j)}$ is the number of calls for district $j$ predicted by the fit and that should be as similar as possible to the real number of calls $N_C^{(j)}$. The number $N_I^{(j)}$ corresponds to those cases whose record was opened in the studied period of time, regardless if the laboratory result was confirmed at some other time.  Observe that proportionality coefficients $\bm \theta = (\theta_P,\theta_I)$ are independent of index $j$ and should be fitted from the data.

There is important information to be extracted once the number of calls for each COVID-19 lab-confirmed case, $ \theta_I$, is fitted.  Let $f_c$ be the fraction of people with symptoms that contacts the Health Care System through the COVID-line, and let $\kappa$ be the average of times each one of these persons calls the COVID-line.  With these two variables we can now estimate the total number of lab-confirmed plus mild cases.  In fact, we can now assert that the number of calls from different persons for each lab-confirmed case is $\theta_I/\kappa$.   Moreover, we can also estimate that for each mild calling the COVID-line, there exist other $1/f_c$ mild cases which are not calling.   Henceforth, we obtain
\begin{equation}
	\mbox{Total number of lab-confirmed plus mild } = \frac{\theta_I / \kappa}{f_c}\, N_I ,
	\label{main}
\end{equation}
where $N_I$ is the number of lab-confirmed cases in the studied district and period of time.  The values for $\kappa$ can be estimated by the telephone records or surveying on the COVID-line reported cases, whereas $f_c$ is usually within the information available from the Health Care Administration.  In any case, the outcome in Eq.~\ref{main} has many sources of intractable systematic uncertainties, and therefore its value should be understood within the corresponding caution.

Finally it is worth discussing a few details concerning the above ideas.  First, observe that mild is not the same as asymptomatic. A mild case is defined as having enough symptoms to place a call to the Health Care System COVID-line, but less than the threshold required to be tested.  Second, notice that, within this framework, the line dividing lab-confirmed and mild cases depends on the local Health Care System policies, since the division comes from the definition of the threshold needed to be tested.  Therefore, the value of the factor accompanying $N_I$  in Eq.~\ref{main} depends on the local Health Care System policies for each region.  At  last, observe that the division between mild and asymptomatic cases, although independent of policies, is rather a smooth division since depends person by person on their perception of the symptoms.  The number that comes out from Eq.~\ref{main} does not consider cases which are purely asymptomatic.


As an example with a real application of the above proposal we study the Buenos Aires Province in Argentina (PBA) during the period of June 2020, in which there were not major changes in weather nor in threshold and methodology for laboratory testing.  During this period of time the number of lab-confirmed cases was in the order of 1000 new cases per day.  The PBA Health Care System runs a COVID-line for symptoms whose local phone number is 148.  The access to this local number consists in an automatic menu that derives into an operator for those cases passing the automatic menu.  We take as $N_C$ the number of calls that choose from the automatic menu the COVID symptoms option.  We count calls even if they hang up before their call is taken by an operator.  Given the structure of this COVID-line, the call counting at this level does not differentiate the district from which the call was placed, therefore we can only take the whole PBA as one district for this specific analysis.  To have many measurements with different $N_I$ --which is key for the fitting-- we take as the period of time of each data point for Eq.~\ref{eq:fit} each full day during June.  Yielding a total of 30 data points.

\begin{figure}[t]
  \begin{center}
\includegraphics[width=0.48\textwidth,height=5cm]{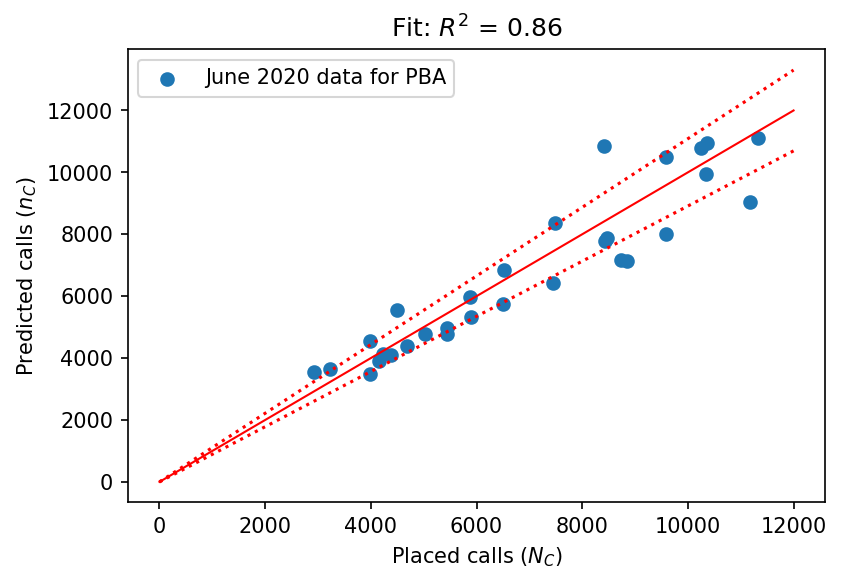}
	  \includegraphics[width=0.48\textwidth, height=5cm]{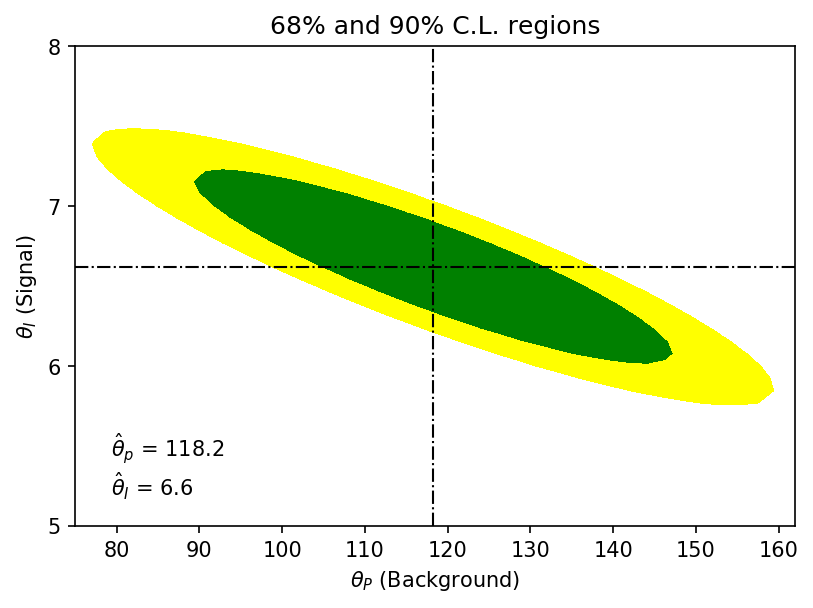}
  \end{center}
	\caption{Left: Placed calls versus predicted calls using the result for the fit from Eq.~\ref{eq:fit} and the population and number of cases for PBA during June 2020.  Each data point corresponds to a day in all PBA.  Solid red line indicates the identity and the dashed red line indicates the average of the fraction of the deviation from identity for all data points, which is 11\%.  Right: Best fit point according to the maximum likelihood and its 68\% and 90\% C.L.~regions (see text).}
  \label{fit}
  \end{figure}

In order to fit the parameters $\bm\theta$ from the data we can use the method of maximum likelihood.  We should maximize the likelihood $L(\bm\theta)$ or equivalently minimize the $\chi^2(\bm \theta)$ defined through
\begin{eqnarray}
	\chi^2(\bm \theta) = -2 \ln L({\bm \theta}) = \sum_j \frac{\left( N_C^{(j)} - ( \theta_P \, N_P^{(j)} + \theta_I \, N_I^{(j)}) \right)^2}{\sigma^{(j)2}}.
	\label{likelihood}
\end{eqnarray}
Observe that also the method of least squares could have been used and results would be similar.  Statistically, the variance $\sigma^{2}$ should correspond to the Poissonian variance of the number of calls, however there are sources of systematic errors which dominate over the statistical one.  In particular, the correspondence between the number of new records opened a given day, and the number of calls that day, yields a systematic uncertainty since both may be shifted one or two days due to intractable causes. More systematic uncertainties may come from other uncontrollable behaviors.  We find that assigning a 11\% systematic uncertainty to the number of calls is self-consistent with the outcome of the fit.  We obtain that the best fit for Eq.~\ref{likelihood} is through
\begin{equation}
	\begin{aligned}
	\hat \theta_P &= 118.2 \mbox{ calls per 1M people} \\
	\hat \theta_I &= 6.6 \mbox{ calls per confirmed case,} 
	\end{aligned}
	\label{thetas}
\end{equation}
and at this point we obtain $\chi^2(\hat{\bm \theta})=\chi^2_{min}=36.0$.  We plot in Fig.~\ref{fit}a the result of this fit by comparing the real number of placed calls against the predicted number of calls coming from inserting the fitted values $\hat{\bm\theta}$ into Eq.~\ref{eq:fit}.  The fit yields a coefficient of determination $R^2 = 0.86$, indicating that the fit is very good, but that there are still extra sources of uncertainties, as it can be seen in the plot. To find the uncertainty in $\hat{\bm \theta}$ we use that the contour in parameter space defined by $\chi^2(\bm\theta) \leq \chi^2_{min} + 2.3$ has a 68\% probability of covering the true value \cite{pdg}.  We plot the resulting 68\% and 90\% confidence level contour regions in Fig.~\ref{fit}b.  If we disregard the value of $\theta_P$, we find $\theta_I = 6.6 \pm 0.4$ calls per confirmed case.

In the above analyzed data for PBA we know from the records that from the total lab-confirmed cases, 22\% correspond to cases that entered into the Health Care System through the COVID-line.  We use this to estimate that the ratio of calls from infected people corresponds to $f_c=0.22$.  On the other hand we have determined through a simple survey that each confirmed person calling the COVID-line makes on average 1.5 calls ($\kappa = 1.5 \pm 0.1$).  Using this into Eq.~\ref{main} we obtain
\begin{equation}
	\mbox{Total number of lab-confirmed plus mild @ PBA} = (20 \pm 2) \, N_I .
	\label{result}
\end{equation}
Where $N_I$ is the total reported cases in PBA.  Observe that in deriving this result there may have been introduced additional systematic sources through the variables $f_c$ and $\kappa$ which have not been taken into account and, in contrast to $\theta_I$, are not controlled by the goodness of fit.  These systematic may consist, for instance, in infected who may have called the COVID-line, but finally entered the system through another path; or in a different ratio of mild than severe cases calling the COVID-line.  The outcome of this algorithm should be understood as an early and inexpensive estimate of the rate of symptomatic to reported COVID-19 cases.  This result should be complemented with serology and/or massive testing results. 

The factor $20\pm 2$ in Eq.~\ref{result} is compatible with results in other parts of the World on the ratio of total infected cases found through serology to reported cases.  For instance, Germany has estimated 10 times more infected than those reported by lab-confirmation \cite{germany}, Spain 15 \cite{spain} and London 45 \cite{london}.  Observe, however, that these last numbers may include the asymptomatic cases, which are beyond the scope of this work and whose role in the COVID-19 Pandemic is still controversial \cite{asymptomatic,diamond,duaa}.

Summarizing, we have designed a novel method to estimate the number of mild cases in the COVID-19 Pandemic.  This method, variations and/or digital adaptations based on the idea in Eq.~\ref{eq:fit}, could be useful as alternatives or complements to massive testings while considerably less expensive.  We use the correlation between the COVID-line phone calls and the number of reported cases and population in each district to estimate how many calls are made for each reported case.  Using in addition the fraction of cases entering the system through the COVID-line and how many time calls are repeated by the same person, we provide an estimation for the total number of reported plus mild cases.  We apply the technique to Buenos Aires Province in Argentina and find numbers compatible with other countries in which serology antibody tests have been performed.

\vskip .5cm
We thank R.~Vaena, E.~Roulet, L.~da Rold, D.~de Florian, M.~Szewc, F.~Lamagna, S.~Crespo, N.~Kreplak and L.~H.~Molinari for useful conversations on the results of this work.  A.E.~thanks Development Bank of Latin America CAF  and CONICET  for partially funding this research.

\bibliography{biblio}
\end{document}